\documentclass[lettersize,journal]{IEEEtran}
\usepackage{amsmath,amsfonts}
\usepackage{algorithmic}
\usepackage{algorithm}
\usepackage{array}
\usepackage{float}
\usepackage{textcomp}
\usepackage{stfloats}
\usepackage{url}
\usepackage{verbatim}
\usepackage{graphicx}
\usepackage{subcaption}
\usepackage{cite}
\usepackage{booktabs}
\usepackage{multirow}
\usepackage{array}

\begin{document}

\title{Robust Reversible Watermarking in Encrypted Images Based on Dual-MSBs Spiral Embedding}

\author{Haoyu Shen,Wen Yin, Zhaoxia Yin\IEEEauthorrefmark{1},Wan-Li Lyu,Xinpeng Zhang
\thanks{Haoyu Shen is with the School of Communication \& Electronic Engineering, East China Normal University, Shanghai 200241, China, also with the Information Technology Center, Shanghai International Studies University, Shanghai 201600, China(email: shy@shisu.edu.cn). }
\thanks{Wen Yin is with the School of Computer Science and Technology, Anhui University, Anhui 230039, China(email: e20301241@stu.ahu.edu.cn).}

\thanks{Zhaoxia Yin is with the School of Communication \& Electronic Engineering, East China Normal University, Shanghai 200241, China(email: zxyin@cee.ecnu.edu.cn).}

\thanks{Wan-Li Lyu is with the School of Computer Science and Technology, Anhui University, Anhui 230039, China(email: lwl@ahu.edu.cn).}

\thanks{Xinpeng Zhang is with the School of Communication and
Information Engineering, Shanghai University, Shanghai 200444, China, also with the School of Computer Science, Fudan University, Shanghai 200411, China(email: zhangxinpeng@fudan.edu.cn).}
}

\markboth{}%
{Shell \MakeLowercase{\textit{et al.}}: A Sample Article Using IEEEtran.cls for IEEE Journals}

\IEEEpubid{}

\maketitle

\begin{abstract}
Robust reversible watermarking in encrypted images (RRWEI) faces an inherent challenge in simultaneously achieving robustness, reversibility, and content privacy under severely constrained embedding capacity. Existing RRWEI schemes often exhibit limited robustness against noise, lossy compression, and cropping attacks due to insufficient redundancy in the encrypted domain. To address this challenge, this paper proposes a novel RRWEI framework that couples dual most significant bit-plane (dual-MSBs) embedding with spatial redundancy and error-correcting coding. By compressing prediction-error bit-planes, sufficient embedding space and auxiliary information for lossless reconstruction are reserved. The dual-MSBs are further reorganized using a spiral embedding strategy to distribute multiple redundant watermark copies across spatially dispersed regions, enhancing robustness against both noise and spatial loss.Experimental results on standard test images demonstrate that the proposed method consistently outperforms under evaluated settings robustness against Gaussian noise, JPEG compression, and diverse cropping attacks, while maintaining perfect reversibility and high embedding capacity. Compared with state-of-the-art RRWEI schemes, the proposed framework achieves substantially lower bit-error rates and more stable performance under a wide range of attack scenarios.
\end{abstract}

\begin{IEEEkeywords}
Dual-MSBs,Spiral,RRWEI,Robustness
\end{IEEEkeywords}

\section{Introduction}
\IEEEPARstart{W}{ith} the continuous evolution of digital imaging technologies and the rapid growth of multimedia data, ensuring content security and copyright protection has become increasingly critical. Reversible data hiding (RDH) \cite{kumar2022reversible}, also known as reversible watermarking, exploits the redundancy of digital media to embed information without introducing permanent distortion to the host. In the absence of attacks, both the original carrier and the embedded data can be perfectly restored, which makes RDH particularly suitable for scenarios requiring high fidelity and security, such as medical imaging \cite{kong2024neural}, military imagery \cite{kumar2023multistage}, and forensic analysis \cite{shi2016reversible}.

Traditional RDH focuses on achieving large embedding capacity while minimizing distortion. A variety of embedding mechanisms have been developed, including difference expansion \cite{tian2003reversible}, prediction error expansion \cite{li2013general}, prediction-error histogram shifting \cite{tsai2009reversible}, and multi-dimensional histogram shifting \cite{fu2021reversible,yamato2014reversible,zhao2018three}. However, watermarks embedded by plaintext-domain RDH are typically fragile; common lossy operations such as JPEG compression or noise may easily destroy the embedded information.

To address this limitation, robust reversible watermarking (RRW)\cite{wang2019independent,gao2023efficient} was introduced, incorporating robustness into reversible embedding. RRW ensures that, when the image under distortions, both the watermark and original image can be perfectly recovered; when the image undergoes moderate distortions, the watermark remains extractable for copyright authentication.

Meanwhile, plaintext-domain watermarking inevitably exposes visual content, posing privacy risks in sensitive fields. Reversible watermarking in encrypted images (RWEI) \cite{puteaux2021survey} integrates image encryption with RDH, enabling data embedding directly in the ciphertext domain. Depending on the embedding order, RWEI methods can be broadly categorized into vacating room after encryption (VRAE) \cite{cheng2025reversible,xiong2023reversible,ke2025two}, Vacating Room By Encryption (VRBE) \cite{fang2025reversible,yu2025reversible,chen2025reversible}and reserving room before encryption (RRBE) \cite{zhang2024multi,ping2024novel,zhang2025reversible,yuan2025reversible,lin2025reversible}. However, most RWEI schemes prioritize embedding capacity while neglecting robustness, making them vulnerable to attacks during transmission.

To overcome this, robust reversible watermarking in encrypted images (RRWEI) has recently emerged, combining ciphertext-domain reversibility with robustness\cite{liang2021robust,xiong2021robust}. Existing RRWEI methods still generally suffer from limited embedding capacity and insufficient robustness, especially under local geometric distortions such as cropping attacks. The fundamental reason is that most current RRWEI schemes are directly migrated from plaintext-domain RRW methods. Since plaintext-domain RRW must strictly control the visual quality of the host image after watermark embedding, its embedding capacity is inherently constrained, and flexible bitstream rearrangement within or across bit planes is typically not permitted. In contrast, watermark embedding in the encrypted domain is free from perceptual constraints and, in principle, offers greater embedding flexibility and more freedom in bit-plane manipulation. However, existing RRWEI methods that rely on migrating plaintext RRW frameworks to the encrypted domain fail to fully exploit this advantage, and instead retain capacity-limited and redundancy-insufficient embedding strategies, which consequently results in poor robustness under cropping attacks. Therefore, merely increasing redundant copies or adjusting embedding locations, without redesigning the ciphertext-domain embedding strategy, is insufficient to significantly improve the robustness of RRWEI against cropping attacks. These limitations motivate the exploration of a new RRWEI framework that can fully exploit the embedding flexibility in the encrypted domain to simultaneously achieve higher robustness, reversibility, and enhanced resistance to geometric attacks.

In response to the above issues, this paper proposes a novel robust reversible watermarking in encrypted images (RRWEI) scheme. The proposed method generates a compressed image using a prediction-error bit-plane compression technique, reserves embedding spaces in the high two bit-planes, embeds the auxiliary information generated during compression into the low bit-planes, and performs bit-level encryption and block shuffling of the image stream. Before embedding, the watermark is processed with adaptive error-correcting coding and then embedded in three copies in a spiral manner across the outer, central, and transition regions of the 8th and 7th bit-planes. At the receiver side, depending on the held keys, it is possible to extract the watermark, recover the image, or perform both simultaneously. Experimental results demonstrate that the proposed scheme achieves high robustness against noise, JPEG compression, and local cropping attacks, while maintaining high embedding capacity and full image reversibility. The main contributions of this work are summarized as follows:
\begin{enumerate}
\item A novel design paradigm for RRWEI is proposed, based on the RWEI framework. Unlike existing schemes that migrate RRW from the plaintext domain, the proposed approach fully leverages the high embedding capacity and flexible embedding regions in the ciphertext domain to optimize robustness while balancing watermark capacity and image reversibility.

\item A dual-MSBs-based embedding strategy coupled with adaptive error-correction coding and spread spectrum modulation is designed. Multiple copies of the watermark are embedded into the reserved spaces of both the 8th and 7th bit-planes, and a majority voting (MV) strategy is applied during extraction to enhance reliability and robustness against noise and compression attacks.

\item A spiral unfolding strategy is tightly coupled with bitstream rearrangement to ensure even distribution of multiple watermark copies across the image. This coupling enhances resistance against local cropping attacks while preserving perfect reversibility of the original image without altering pixel values.
\end{enumerate}

\section{related work}

\subsection{Robust Reversible Watermarking in the Plaintext Domain}
In this work, we select three representative RRW methods and adapt them to the encrypted domain. The following subsection provides an overview of these approaches.
Xiao et al. \cite{xiao2025robust} proposed a robust reversible watermarking scheme based on histogram shifting of DCT coefficients. The image is divided into blocks and transformed using block-wise DCT. The (1,2) and (1,3) coefficients are gathered to form coefficient matrices whose histograms are modified for reversible and robust embedding.

Wang et al. \cite{wang2024robust} introduced a scheme that resists cropping attacks by embedding the same watermark into multiple circular regions determined by adaptive feature-point detection. Reversible watermarking is applied outside these regions, providing robustness against local geometric distortions.

Tang et al. \cite{tang2024robust} developed a robust reversible watermarking framework using polar harmonic transform (PHT) moments. The robust watermark is embedded into moments within the inscribed circle using multi-level quantization spread spectrum modulation, while auxiliary information is reversibly stored outside the circle.

In recent years, learning-based robust watermarking methods have attracted increasing attention due to their strong robustness against noise and watermark removal attacks.
Guo et al.\cite{guo2025robust} proposed a variational autoencoder-based RRW framework that achieves imperceptible distortion and demonstrates resistance to neural watermark removal.
Kou et al.\cite{kou2025iwrn} introduced a deep neural network for blind robust watermarking of artwork images, exhibiting strong robustness against noise attacks. However, these learning-based approaches are inherently designed for the plaintext domain and cannot be directly migrated to the encrypted domain.
Their effectiveness relies on semantic consistency, differentiable pixel intensities, and feature learning from natural images. In contrast, ciphertext images generated by secure encryption schemes exhibit near-uniform distributions and lack semantic structures, rendering neural feature extraction and gradient-based optimization ineffective.
Therefore, although learning-based watermarking methods demonstrate promising performance in the plaintext domain, they are not suitable for RRWEI.

\subsection{Robust Reversible Watermarking in Encrypted Images (RRWEI)}
At present, there are two representative works in the RRWEI field.Liang et al. \cite{liang2021robust} presented an RRWEI method based on the Paillier cryptosystem. The image is encrypted block-wise, and statistical features of each block are computed. Using the homomorphic property of Paillier encryption, histogram shifting is performed directly in the ciphertext domain, allowing watermark extraction from either encrypted or decrypted images. Xiong et al. \cite{xiong2021robust} proposed an RRWEI scheme built on secure multi-party computation (SMC). The image is encrypted using additive secret sharing and block permutation. A two-stage RRWEI framework is adopted, combining Patchwork-based robust watermarking with prediction error expansion (PEE)-based reversible embedding. The scheme also incorporates Kronecker compressed sensing to reduce computational complexity during encoding.

Although the above methods provide robustness or ciphertext-domain embedding, they generally inherit constraints from plaintext-domain RRW frameworks. The embedding capacity is limited, redundancy is insufficient, and most schemes embed only a single copy of the watermark. As a result, the watermark cannot be reliably extracted under strong noise or geometric attacks such as cropping, highlighting the need for more robust RRWEI solutions.

\section{proposed method}
\begin{figure*}[htbp]
     \centering
    \includegraphics[width=0.8\textwidth]{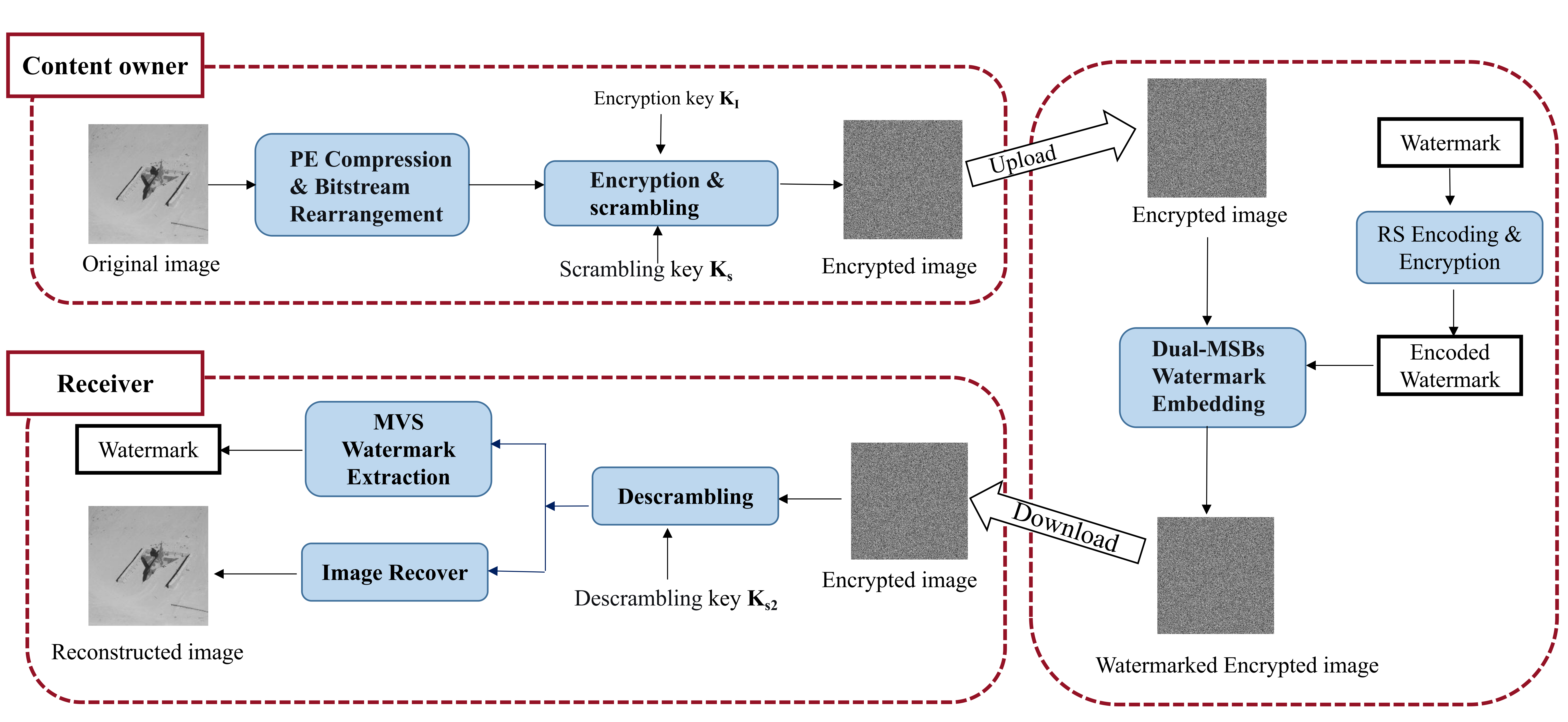} 
    \caption{The overall diagram of the proposed method}
   \label{fig:framework}
\end{figure*}

Motivated by the above observations, this work explicitly redesigns the ciphertext-domain embedding strategy by exploiting bit-plane rearrangement and redundancy-aware distribution, rather than directly migrating plaintext-domain RRW mechanisms. The overall framework of this paper is shown in Fig.\ref{fig:framework}, which consists of five stages: reserving embedding space, rearranging the dual-MSBs, image encryption, watermark embedding, watermark extraction, and image restoration.

\subsection{Reserving Embedding Space} 
This scheme adopts the prediction error bit-plane compression technique proposed by Yin et al. \cite{yin2020reversible} to generate the compressed image. During the compression process, auxiliary information for image restoration is generated, including: block size (4 bits), threshold T (3 bits), and the quantity and positions of overflow pixels (each overflow pixel occupies 18 bits for storage). The auxiliary information is stored in the Least Significant Bits (LSBs) plane of the compressed image. After compression, the bitstream of the image is arranged sequentially from the lower bit planes to the higher bit planes, immediately following the auxiliary information.
\subsection{Rearranging the Dual-MSBs} 
This scheme embeds the same watermark in the same regions of the dual-MSBs using spread spectrum embedding. Therefore, it is necessary to reserve space for watermark embedding in the dual-MSBs. Based on the compression results, the maximum embedding capacity for the watermark is calculated and sent as intermediate temporary auxiliary information to the watermark embedder. This prevents damage to the image bitstream caused by embedding watermarks that are too long. When the net embedding space obtained from compression is greater than or equal to 2 bits per pixel(bpp), the dual-MSBs do not require rearrangement of the bitstream to complete watermark embedding. In this case, only the watermark embedding length needs to be stored in the three reserved watermark length embedding regions of the 8th and 7th bit planes, and the rearrangement flag is set to 0, indicating no rearrangement.
When the net embedding space obtained from compression is less than 2 bpp, rearrangement of the watermark bitstream is required. To enhance robustness against cropping attacks, this paper adopts a spiral traversal method to unfold the two-dimensional image into a one-dimensional bitstream, i.e., traversing pixels from the outer edges of the image layer by layer toward the center. The specific rearrangement method is shown in Fig.\ref{embeddingdiagram}. Two bitstream embedding regions are reserved in the middle of the three watermark copies for the rearranged watermark bitstream. The arrangement of the watermark bitstream starts from the bitstream embedding region in the 7th bit plane. After the reserved embedding region in the 7th bit plane is filled, the embedding continues in the reserved embedding region of the 8th bit plane until completion. In this case, the rearrangement flag must be set to 1, indicating that rearrangement has been performed. Upon receiving this flag, the receiver must perform a restoration rearrangement operation during image recovery. Additionally, to enable the receiver to losslessly restore the image when no attacks occur, the length of the rearranged bitstream must be embedded after the watermark capacity and rearrangement flag. This rearranged bitstream length is divided into two parts and embedded after the first watermark capacity and before the last watermark capacity, respectively.

\subsection{Image Encryption} 
After the image bitstream rearrangement and auxiliary information processing, bit-level encryption is performed on the image. Then, image shuffling is carried out using the image shuffling key to further enhance the security of the encrypted image.

\subsection{Watermark Embedding}
Before embedding the watermark, error correction coding(ECC) is applied to enhance its robustness, which can correct errors introduced during transmission due to attacks such as noise or JPEG compression. This paper employs the Reed-Solomon (RS) code as the ECC. Specifically, the adopted coding scheme is RS(31, 3), which constructs codewords over the Galois field $\operatorname{GF}(2^5)$. Each codeword has a length of n = 31 symbols, containing k = 3 symbols of effective watermark information, with each symbol represented by a 5-bit binary number. Its error correction capability is expressed by Formula \ref{eq:error_correction}:
\begin{equation}
t = \frac{n - k}{2}
\label{eq:error_correction}
\end{equation}

Here, $t$ represents the Error Correcting Ability , which means the number of erroneous symbols a codeword can correct. For the RS(31, 3) code adopted in this paper, 
$t=14$ can be derived. This means that each codeword, consisting of 31 symbols ( 155 bits in total), can correct up to 14 arbitrary symbols (up to 70 bits) of error. $n$ denotes the codeword length (in symbols), and $k$ denotes the length of watermark information symbols within a codeword. From Formula \ref{eq:error_correction}, it can be inferred that the fewer the watermark information symbols $k$ embedded in a codeword, the stronger its error correction ability $t$, but the lower the corresponding coding efficiency .The parameters selected in this paper strike a balance between error correction strength and embedding capacity.

\begin{figure}[htbp]
\centering
\includegraphics[width=0.4\textwidth]{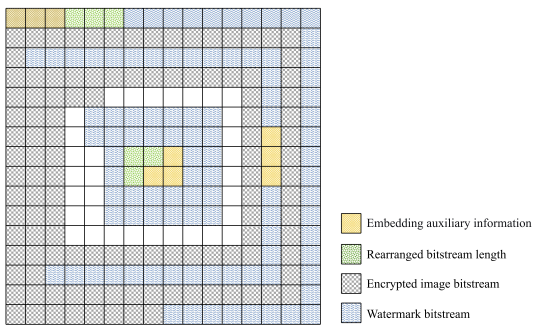} 
\caption{Diagram of Watermark Spread Spectrum Embedding}
\label{embeddingdiagram}
\end{figure}

As shown in Fig. \ref{embeddingdiagram}, to enhance the robustness of the watermark, this scheme employs spread spectrum technology to embed three copies of the watermark in each of the 8th and 7th bit planes of the image, distributed across the outer, central, and intermediate transition regions of the respective bit planes. Upon receiving the encrypted image processed by the image owner, the watermark embedder first extracts the watermark capacity information and the rearrangement flag to determine the embedding regions. Subsequently, the actual watermark length information is written into the originally reserved capacity information positions, followed immediately by the spread spectrum embedding of the rearrangement flag. Finally, using spread spectrum embedding, the six copies of the watermark are respectively embedded into the 8th bit plane (three copies) and the 7th bit plane (three copies). All embedding operations follow the same spiral sequence as the rearrangement stage, ensuring precise placement of the watermark into the pre-allocated spatial positions.

\subsection{Watermark Extraction and Image Restoration}
Based on the keys in their possession, the receiver can extract the watermark, restore the image, or perform both tasks.

Case 1: If the receiver only possesses the watermark decryption key, they first extract the three copies of the watermark length information from the 8th bit plane, along with the rearrangement flag located after the watermark length information. This determines the positions for extracting the three watermark copies. Next, the receiver uses image morphological operations to determine whether the watermark has been subjected to cropping attacks. If a cropping attack is detected, a cropping location map is generated. Based on this map, the receiver identifies which of the three extracted encoded watermark bits are unaffected by cropping and uses them as the final result. If no cropping attack is detected, a MV strategy is applied to the three watermark copies to determine the final extracted watermark value. The specific watermark extraction method is shown in Formula \ref{eq:watermark_extraction}:

{\small
\begin{equation}
w = 
\begin{cases} 
1, & \text{if } \sum_{i=1}^3 s_i = 3 \text{ and } \sum_{i=1}^3 b_i \ge 2 \\
0, & \text{if } \sum_{i=1}^3 s_i = 3 \text{ and } \sum_{i=1}^3 b_i < 2 \\
b_k, & \text{if } 1 \le \sum_{i=1}^3 s_i \le 2, \text{ where } k = \min\{i \mid s_i = 1\} \\
0, & \text{if } \sum_{i=1}^3 s_i = 0
\end{cases}
\label{eq:watermark_extraction}
\end{equation}
}
Here, $w$ is the determined final watermark bit, $b1,b2,b3$ are the corresponding bits from the three watermark copies, and $S1,S2,S3$ are the corresponding bits from the cropping location map.

Case 2: If the receiver only possesses the image decryption key, they first extract the three copies of the watermark length information from the 8th bit plane, along with the rearrangement flag located after the watermark length information. Then, the image bitstream is decrypted. If the rearrangement flag is set to 1, a restoration rearrangement operation is required. The rearranged watermark length information is extracted, followed by the extraction of the rearranged bitstream. The bitstream is then restored to its original order, i.e., sequentially arranged starting from the top-left corner of the 7th bit plane. Finally, the image bitstream is decompressed to restore the corresponding bit planes, thereby reconstructing the original image.

\section{experiment}
In this section, we first experimentally evaluate the impact of the two strategies proposed in this paper on watermark robustness. The selected attacks include three common types encountered by images in communication channels: additive noise, lossy compression, and cropping attacks. We then compare the proposed scheme with five existing robust reversible watermarking schemes for encrypted images. Among them, Liang et al. \cite{liang2021robust} and Xiong et al. \cite{xiong2021robust} are RRWEI schemes, while Xiao et al.\cite{xiao2025robust}, Wang et al. \cite{wang2024robust}, and Tang et al. \cite{tang2024robust} are robust reversible watermarking schemes designed for plaintext images; these three schemes are migrated to the ciphertext domain for comparison. We emphasize that all plaintext-domain methods are migrated in a straightforward manner without altering their core embedding mechanisms. The computational platform for this experiment is a server running the Windows Server 2016 operating system, logically configured with 48 CPU cores and 128 GB of RAM. All algorithm implementation, simulation, and data processing were conducted in the MATLAB R2022a environment.

\subsection{Robustness–Redundancy Trade off Analysis}
\begin{figure}[htbp]
\centering
\includegraphics[ width=0.45\textwidth]{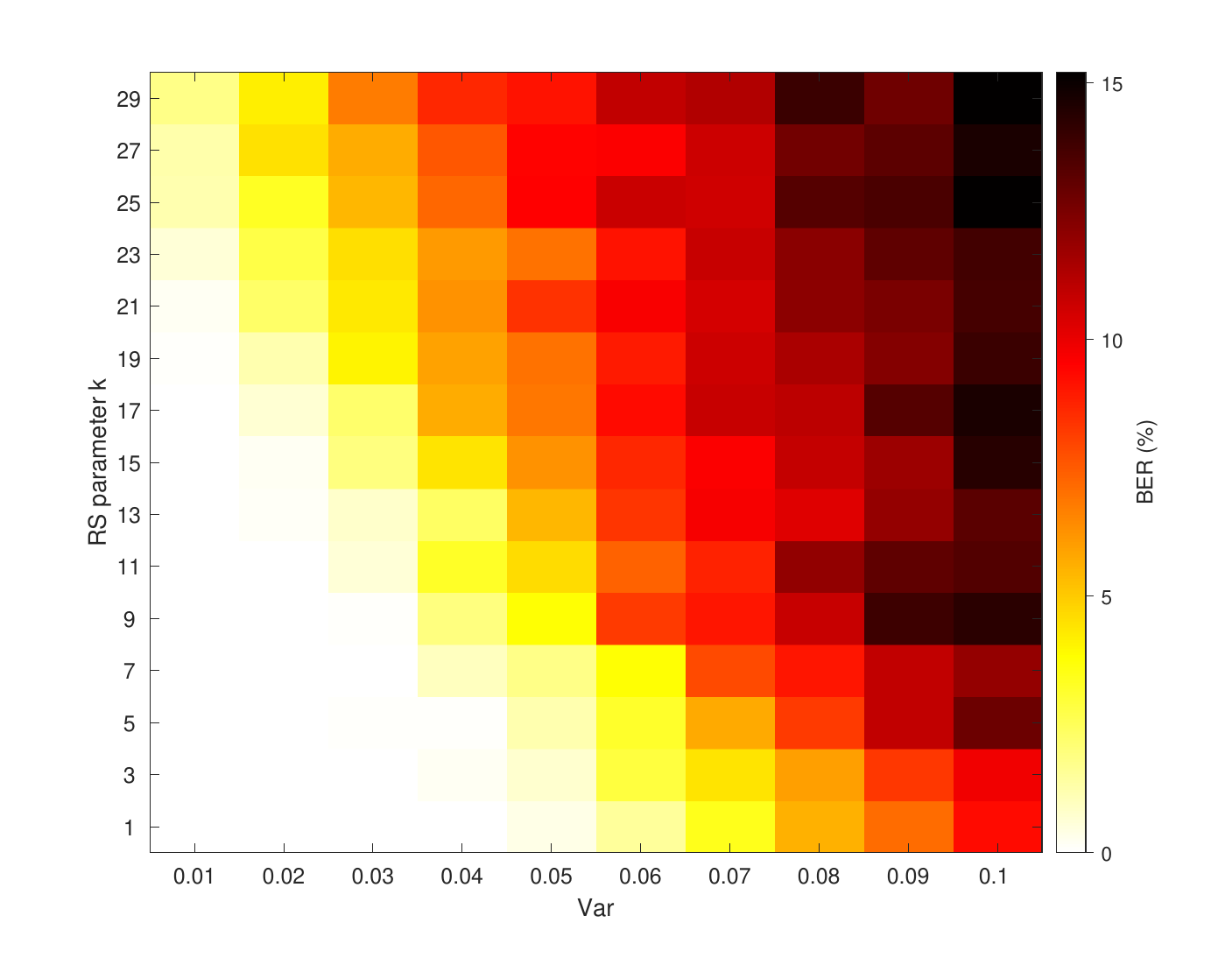} 
\caption{Average BER under Different RS Parameters and Noise Variances }
\label{fig:heat}
\end{figure}

Fig.\ref{fig:heat} illustrates the BER performance under different RS redundancy settings (parameter k) and Gaussian noise variances. The experiments were conducted on six commonly used images: $Baboon$, $Airplane$, $Lake$, $Pepper$, $Boat$ and $Barbara$.The RS parameter $k$ ranging from 1 to 29 and the noise variance varying from 0.01 to 0.1. As shown in Fig.\ref{fig:heat}, stronger redundancy (smaller k) significantly reduces BER, especially under moderate-to-high noise conditions. However, this comes at the cost of increased coding overhead and reduced effective payload.
This observation motivates the choice of a medium-strength RS(31,3) code combined with dual-MSBs spatial redundancy, achieving a favorable robustness–capacity trade-off.

\subsection{Robustness-oriented Analytical Intuition under Simplified Assumptions}
The following analysis adopts simplified statistical assumptions to provide analytical intuition into the relative noise robustness of different embedding strategies in the encrypted domain. This analysis provides an intuitive explanation rather than a formal proof.

Under a well-designed encryption scheme, ciphertext pixel values are widely approximated as being uniformly distributed in order to resist first-order statistical attacks. This assumption is empirically supported by the histogram and entropy analysis reported in Section~\ref{sec:stat_security}, where encrypted and watermarked images exhibit nearly uniform pixel distributions. Based on this commonly adopted abstraction, we analyze the relative tendency of watermark bit flipping under additive Gaussian noise for MSB-only and dual-MSBs embedding strategies. 

It should be emphasized that the dual-MSBs strategy intentionally introduces a dependency between the two highest bit-planes to enhance robustness.

In the MSB-plane embedding scheme, the watermark pixel values can be modeled as uniformly distributed across the full pixel range, i.e., $X_1 \sim U(0,255)$. To compare the robustness of the proposed dual-MSBs embedding scheme with the MSB-only scheme under Gaussian noise, we use the \textit{bit-flip probability} to quantify the likelihood of a single watermark bit being flipped under different noise variances. Let the additive noise $n \sim \mathcal{N}(0, \sigma^2)$. 

For watermark pixels in $[0,127]$, the flipping condition is $x+n \ge 127.5$ (the final pixel value is rounded to the nearest integer and clipped to $[0,255]$). For a given pixel value $x$, this requires $n \ge 127.5-x$, and the bit-flip probability is

\begin{equation}
P_\text{flip} = 1 - \Phi\left(\frac{127.5 - x}{\sigma}\right),
\end{equation}

where $\Phi$ is the cumulative distribution function (CDF) of the standard normal distribution. Similarly, for pixels in $[128,255]$, the flipping condition is $x+n \le 127.5$, and the corresponding flip probability is

\begin{equation}
P_\text{flip} = \Phi\left(\frac{127.5 - x}{\sigma}\right).
\end{equation}

Thus, the overall bit-flip probability for the MSB-plane embedding scheme can be written as:
{\tiny
\begin{equation}
F_{\text{flip}}=\frac{1}{256}
\left[
\sum_{x=0}^{127}\left(1-\Phi\!\left(\frac{127.5-x}{\sigma}\right)\right)
+\sum_{x=128}^{255}\Phi\!\left(\frac{127.5-x}{\sigma}\right)
\right]
\label{0127}
\end{equation}
}

 However, in the dual-MSB embedding scheme, since the two highest bits are restricted to either 00 or 11, the watermarked ciphertext pixel values are intentionally constrained to be approximately uniformly distributed within the intervals [0,63] and [192,255]. Thus, they can be modeled as $X_2 \sim U([0,63]\cup[192,255])$.

Therefore, the bit-flip probability for a single watermark bit under the dual-MSBs embedding scheme is:
{\tiny
\begin{equation}
F_{\text{flip}}=\frac{1}{128}
\left[
\sum_{x=0}^{63}\left(1-\Phi\!\left(\frac{127.5-x}{\sigma}\right)\right)
+
\sum_{x=192}^{255}\Phi\!\left(\frac{127.5-x}{\sigma}\right)
\right]
\label{128255}
\end{equation}
}

\begin{figure}[htbp]
\centering
\includegraphics[ width=0.45\textwidth]{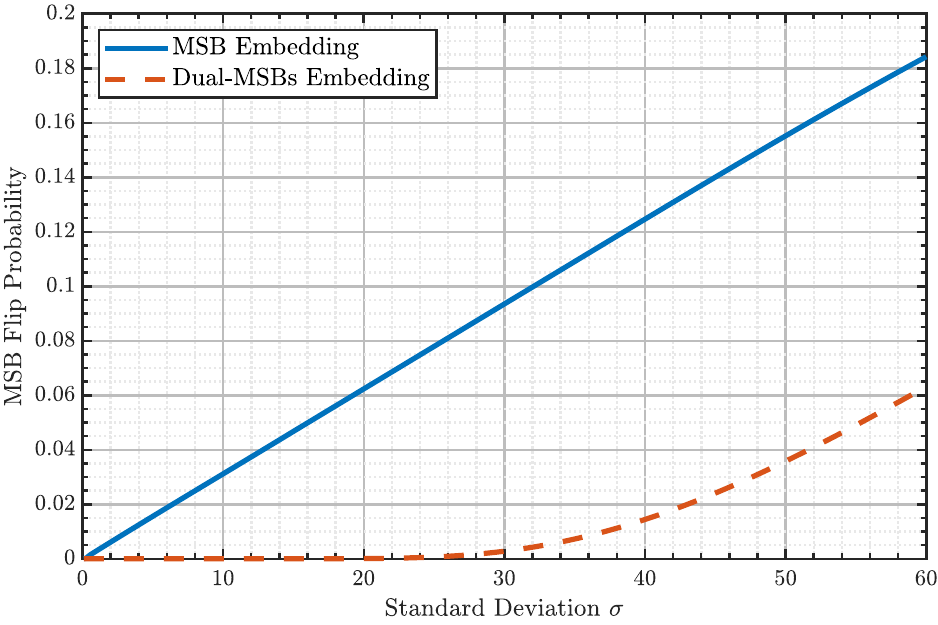} 
\caption{Bit-flip probability versus Gaussian noise standard deviation for MSB-plane and dual-MSBs embedding schemes. }
\label{fig:bitflip}
\end{figure}

As shown in Fig.~\ref{fig:bitflip}, embedding redundancy across dual most significant bit-planes significantly reduces the probability of watermark bit flipping under additive noise. Compared with embedding solely in a single MSB, the proposed dual-MSBs strategy effectively enlarges the valid value interval associated with each embedded bit, thereby increasing its tolerance to noise-induced perturbations. As the noise variance increases, this advantage becomes more pronounced, which provides analytical intuition for the superior BER performance observed under severe noise and compression attacks.

It is worth noting that the above analysis considers individual bit-flip behavior only; in practice, the overall BER is further reduced by error-correcting coding and MV, as confirmed by the experimental results.

\subsection{Ablation Study on Watermark Robustness}
To clearly demonstrate the effectiveness and rationality of the proposed design choices, this section presents ablation studies on the Dual-MSBs embedding and spiral embedding strategies, with the goal of isolating and evaluating their individual contributions to the performance of the proposed method.

\begin{figure*}[htbp]
\centering
\begin{subfigure}[t]{0.48\textwidth}
    \centering
    \includegraphics[width=\linewidth]{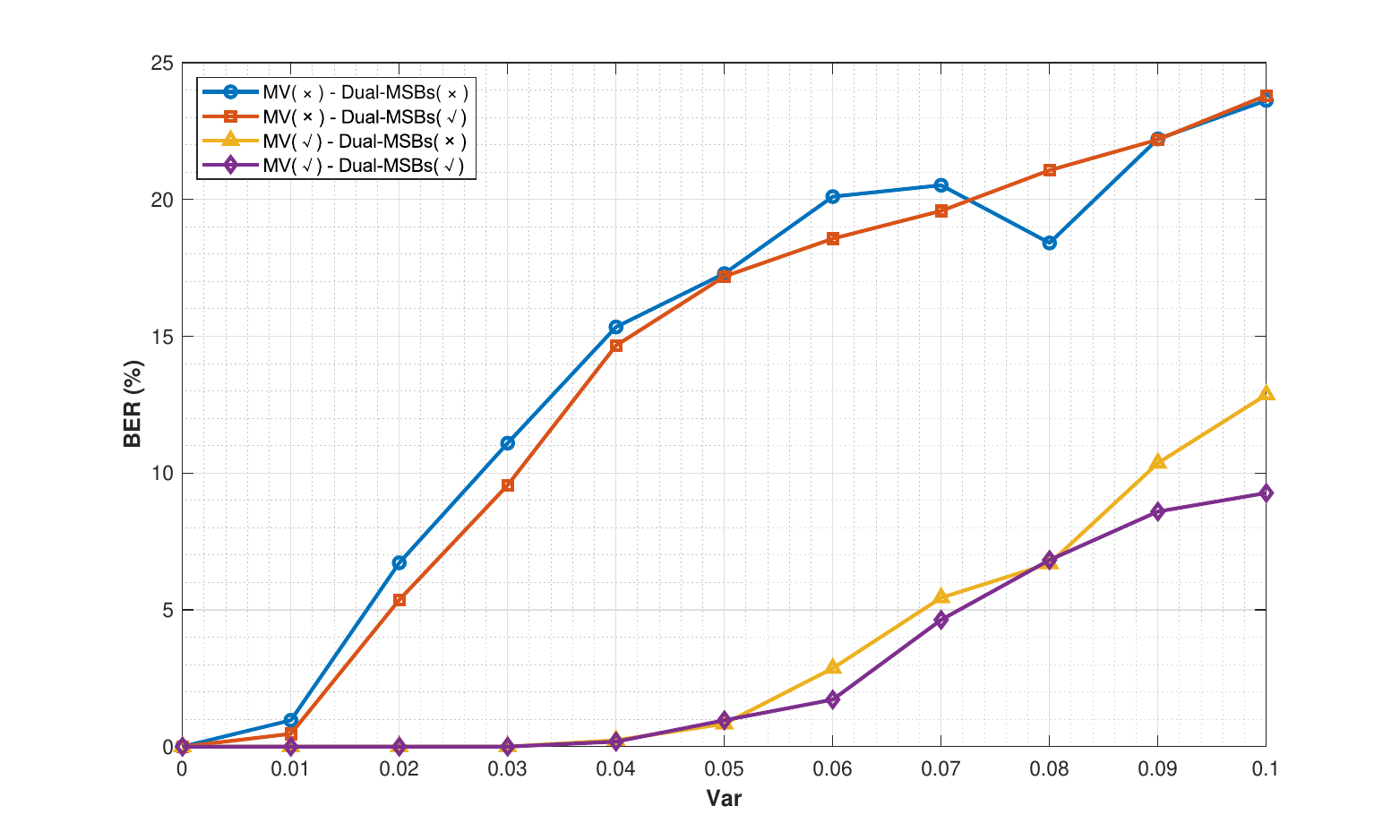}
    \caption{}
    \label{fig:gauab1}
\end{subfigure}
\hfill
\begin{subfigure}[t]{0.48\textwidth}
    \centering
    \includegraphics[width=\linewidth]{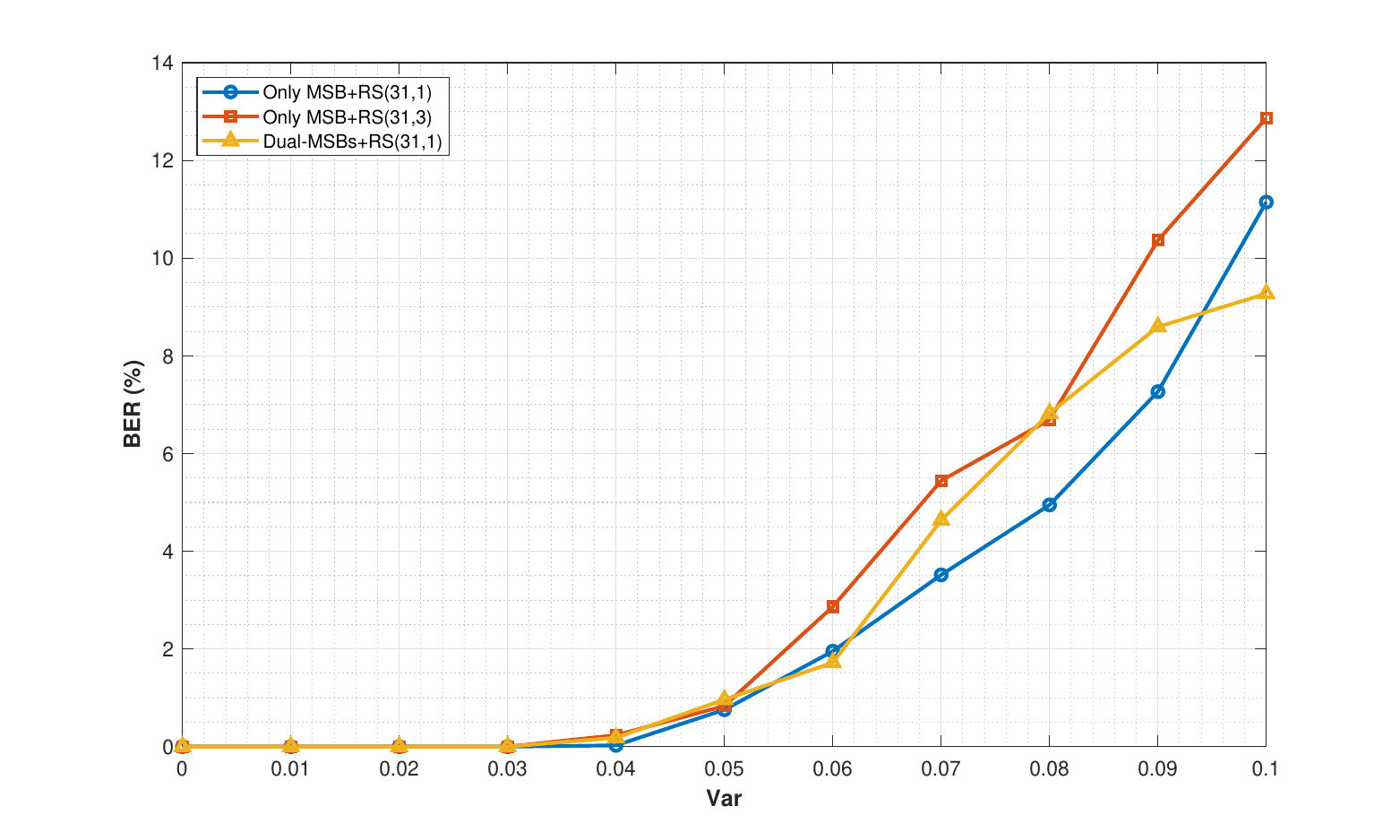}
    \caption{}
    \label{fig:gauab2}
\end{subfigure}
\caption{Robustness under Gaussian noise attacks with different component ablations ($EC=128$ bits).}
\label{fig:gau_ablation}
\end{figure*}

Fig. \ref{fig:gauab1} investigates the robustness behavior of different component combinations under Gaussian noise, with the objective of examining how the dual-MSBs embedding and majority voting (MV) jointly influence watermark extraction reliability.

It should be emphasized that, in the encrypted domain, these components are not strictly independent. In particular, the dual-MSBs embedding enlarges the decision margin at the bit level, while MV further aggregates multiple noisy observations. Therefore, the observed robustness gain reflects the combined effect of bit-level redundancy and decision-level aggregation rather than the contribution of a single isolated factor.

As shown in Fig. \ref{fig:gauab1}, when MV is not applied, the dual-MSBs embedding alone exhibits a moderate reduction in BER in the medium-to-high noise range compared with single-MSB embedding. After incorporating MV, the BER is further reduced for both embedding strategies. Notably, under identical voting conditions, the dual-MSBs-based scheme consistently achieves lower BER across the evaluated noise range, indicating that the dual-MSBs embedding provides a more favorable basis for majority voting under noisy perturbations.

These results suggest that the robustness improvement should not be attributed solely to either dual-MSBs embedding or majority voting. Instead, the dual-MSBs strategy enhances robustness primarily by amplifying the effectiveness of MV through an enlarged bit-level decision margin, leading to more reliable aggregation under noise.

Fig.\ref{fig:gauab2} compares the proposed dual most significant bits (Dual-MSBs) embedding strategy combined with a medium-strength RS(31,3) code against conventional MSB-based schemes employing different error-correcting codes. As expected, when extremely strong channel coding such as RS(31,1) is adopted, the robustness gain introduced by the error-correcting code itself dominates the overall performance, resulting in lower bit error rates under high noise levels.

This comparison, however, also highlights a robustness–efficiency trade-off under the evaluated experimental settings. Specifically, the Only-MSB + RS(31,1) scheme attains improved robustness in high-noise conditions at the cost of substantial coding redundancy introduced by the strong error-correcting code, in addition to embedding three watermark replicas. In contrast, the proposed Dual-MSBs scheme introduces additional spatial redundancy at the embedding stage by employing six watermark replicas, which effectively enlarges the bit-level decision margin of each embedded watermark bit.

Consequently, by combining Dual-MSBs embedding with a medium-strength RS(31,3) code, comparable robustness can be achieved in low-to-moderate noise regimes without resorting to extremely strong channel coding. These results suggest that the proposed Dual-MSBs mechanism is not intended to replace very strong error-correcting codes, but rather to improve robustness efficiency by trading off spatial redundancy and coding redundancy under constraints of limited embedding capacity and a moderate coding budget. This characteristic is particularly relevant for reversible watermarking scenarios, where a balanced compromise between controllable embedding overhead and robustness performance is generally preferred over pursuing the minimum achievable error rate under extreme noise conditions.

\begin{figure}[htbp]
\centering
\includegraphics[ width=0.45\textwidth]{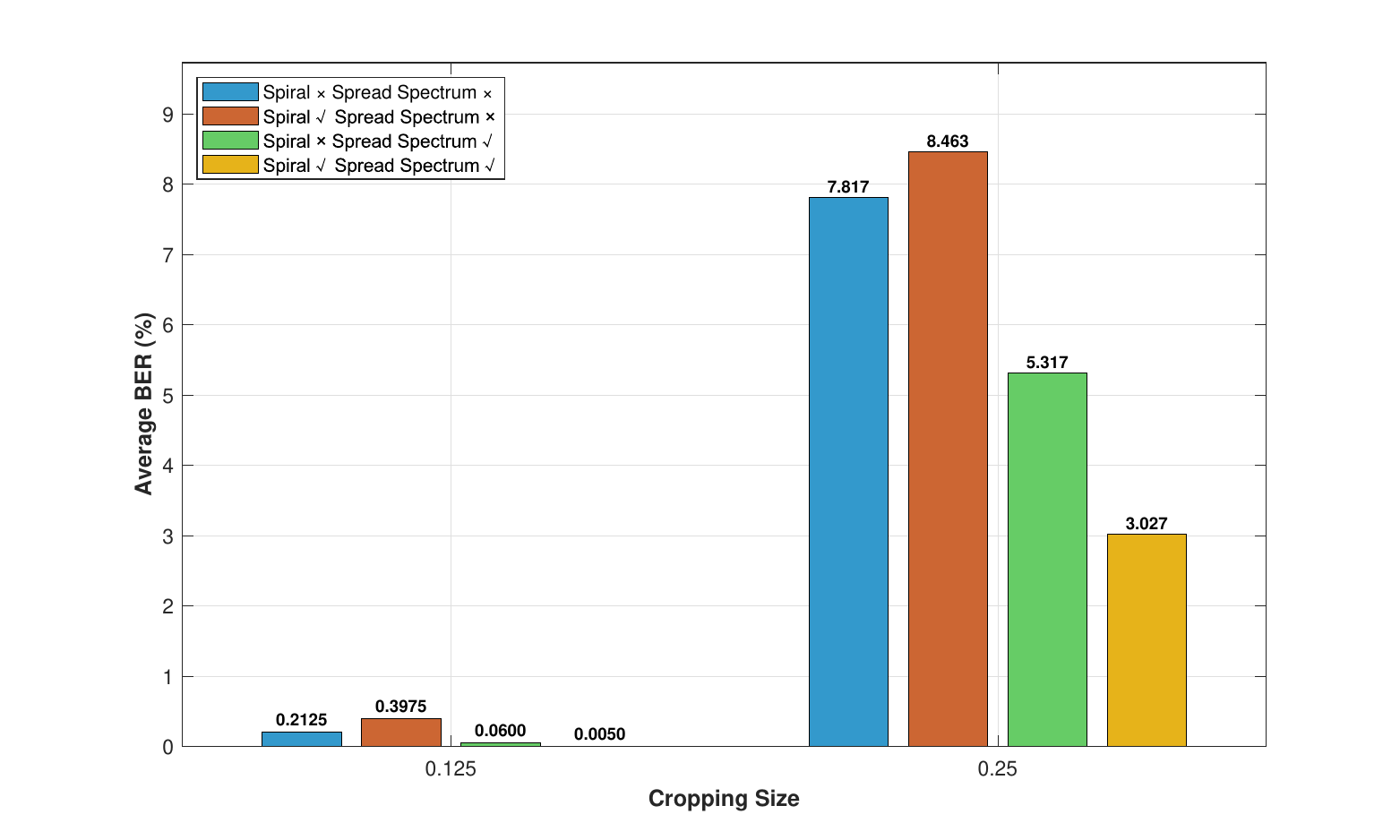}
\caption{Robustness under crop Attacks with Different Component Ablations(when EC=128bits)}
\label{crop}
\end{figure}

This study evaluates the impact of our proposed robustness improvements on resistance against cropping attacks through controlled variable testing. Since the dual-MSBs embedding operates on the same positions in the two most significant bit planes and the receiver extracts watermarks only from the most significant bit plane, this strategy does not influence robustness against cropping attacks. Therefore, Fig.\ref{crop} examines only the effects of spiral embedding and spread spectrum embedding. The experiments are divided into four scenarios based on the adoption of these two strategies. For spiral embedding, the control group uses sequential embedding. For spread spectrum embedding, the control group extracts only one of the three embedded watermark copies.

Both sequential embedding and spiral embedding benefit from the application of spread spectrum embedding, as evidenced by a reduction in the bit error rate. Furthermore, under the condition of using spread spectrum embedding, spiral embedding achieves a lower average bit error rate compared to sequential embedding across various cropping positions and sizes. This demonstrates that spiral embedding makes it less likely for the same bits across the three watermark copies from spread spectrum embedding to be simultaneously cropped out. Consequently, spiral embedding enhances the robustness of the watermark against cropping attacks.

\subsection{Robustness of the Watermark}
\begin{figure*}[htbp]
\centering 
\begin{subfigure}[t]{0.45\textwidth} 
    \includegraphics[width=\linewidth]{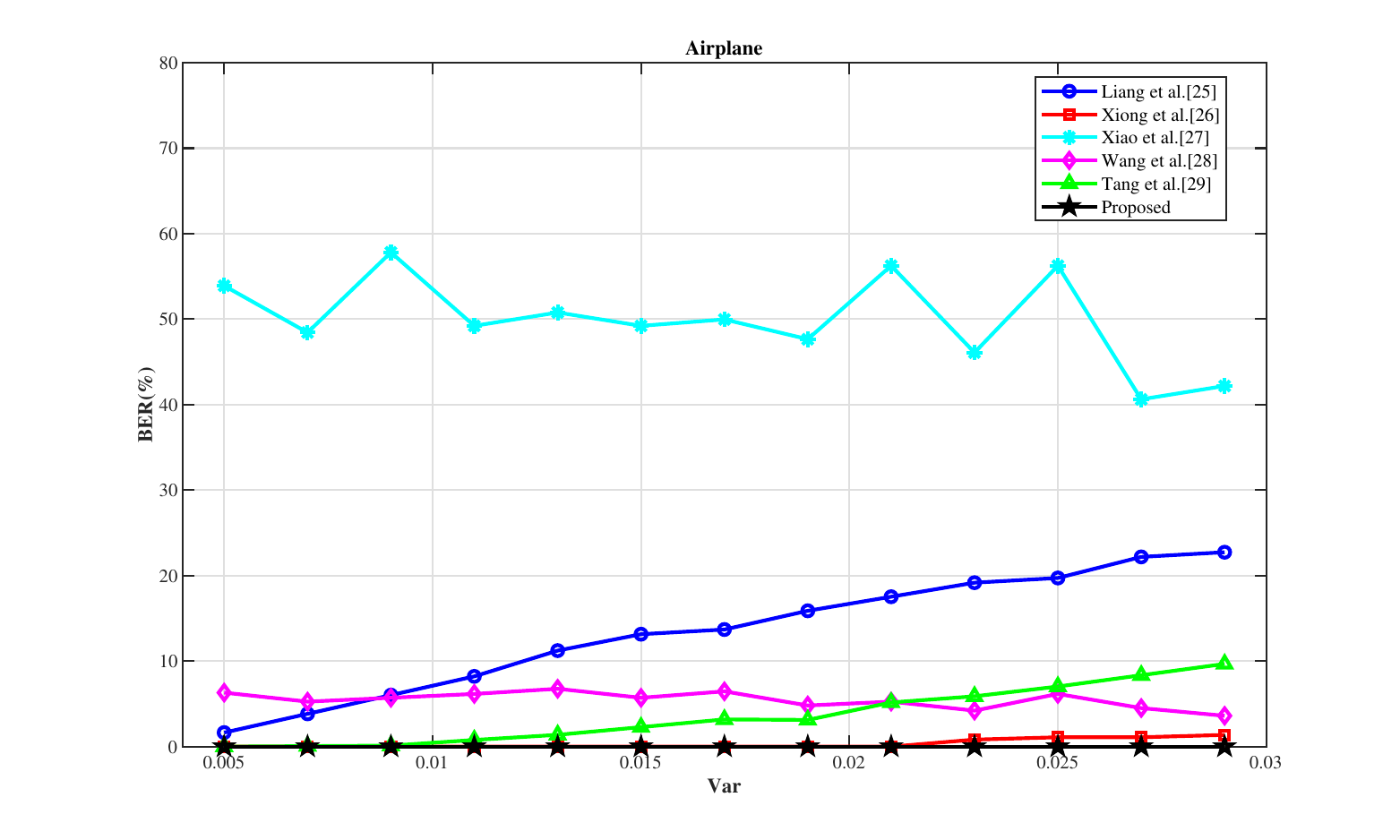} 
    \label{fig:airplane-gaussian}
\end{subfigure}
\hfill 
\begin{subfigure}[t]{0.45\textwidth}
    \includegraphics[width=\linewidth]{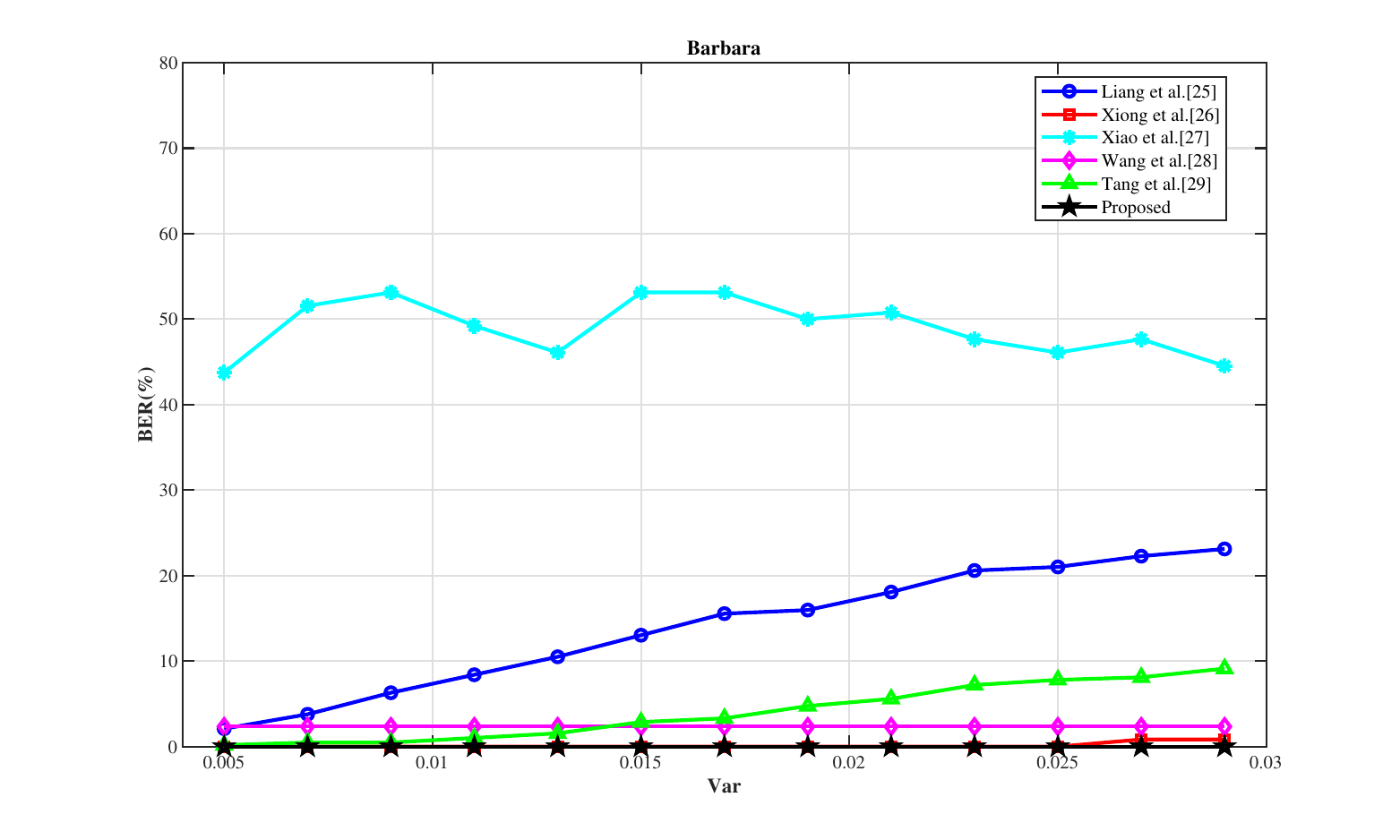}
    \label{fig:barbara-gaussian}
\end{subfigure}
\caption{Comparison of Watermarking Robustness against Gaussian Noise on Airplane and Barbara Images(when EC=128bits)}
\label{fig:gaussian-comparison}
\end{figure*}

First, the robustness of different encrypted-image-based reversible watermarking schemes against additive noise was evaluated, with Gaussian noise chosen as a representative of additive noise. As shown in Fig. \ref{fig:gaussian-comparison}, Xiao et al.\cite{xiao2025robust} relies on a homomorphic encryption scheme, so the addition of noise disrupts the homomorphic decryption at the receiver, resulting in the watermark being completely unrecoverable. In this paper, the proposed scheme achieves a zero bit error rate on both images when the Gaussian noise variance is less than 0.03, outperforming the other five comparison schemes.

To evaluate the performance of the proposed scheme against other RRWEI schemes on a dataset, a test set of 100 images was randomly selected from the BOSSbase dataset\cite{bas2011break}. The Wang et al.\cite{wang2024robust} scheme, which embeds multiple copies of the watermark in circular regions centered at multiple selected feature points to achieve resistance against cropping, fails on extremely smooth images such as “4990.pgm” in the test set because multiple feature points cannot be located. As a result, the embedding cannot proceed, and the bit error rate (BER) reported for this scheme in the table is the average BER computed over the remaining 99 images.

\begin{figure}[htbp]
\centering
\includegraphics[width=0.45\textwidth]{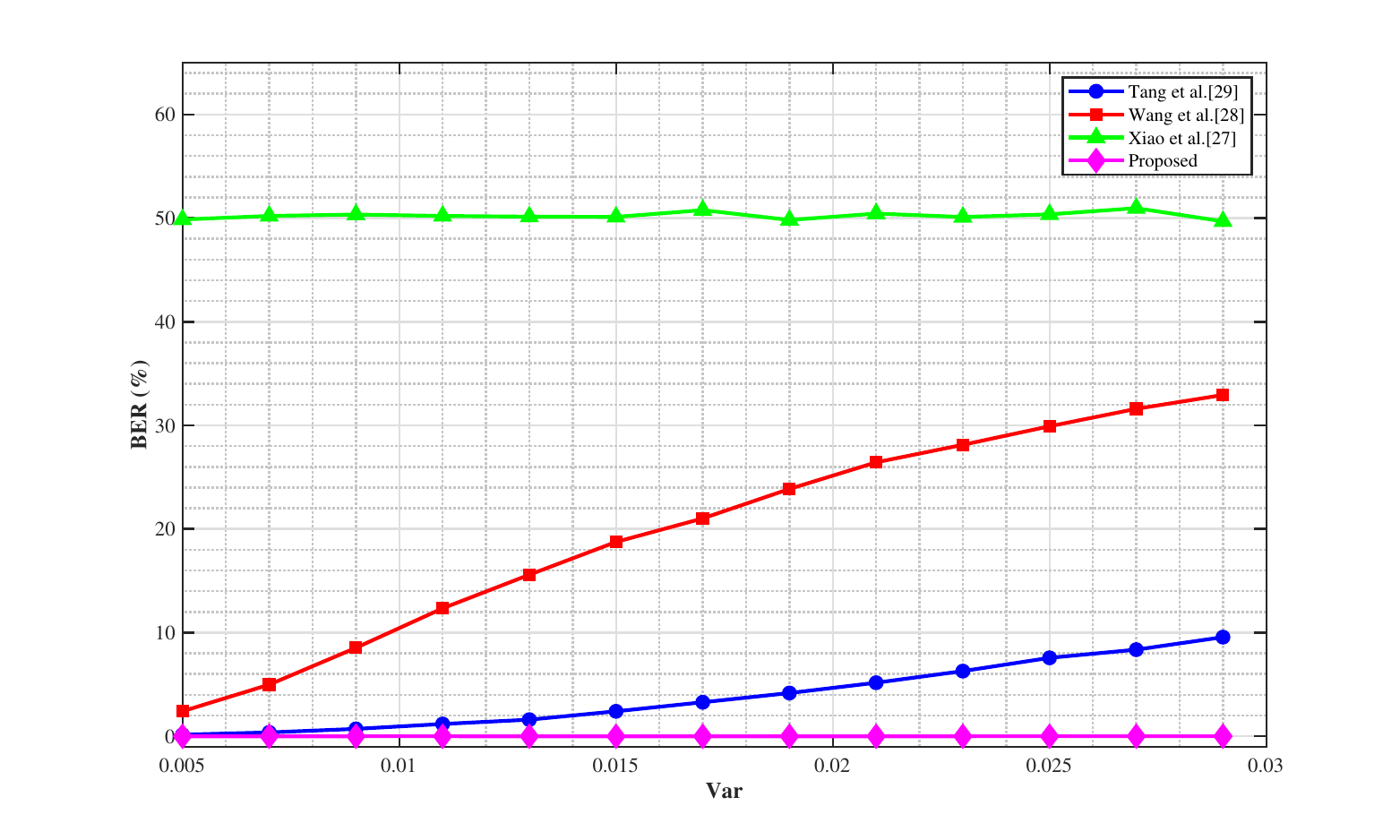} 
\caption{BER Performance Under Gaussian Noise Attack on BOSSbase Dataset(when EC=128bits)}
\label{bossgau}
\end{figure}

Fig.\ref{bossgau} compares the robustness against Gaussian noise of the proposed scheme and three other RRWEI schemes on the test set of 100 randomly selected BOSSbase images. It can be observed that, in terms of average BER over a large number of images, the proposed scheme still achieves the lowest BER.

Table \ref{fixed crop} presents the robustness test for fixed-position cropping attacks. The test set consisted of 100 randomly selected test images from the BOSSbase dataset. The proposed scheme was compared with three other schemes,Tang et al.\cite{tang2024robust}, Wang et al.\cite{wang2024robust}, and Xiao et al.\cite{xiao2025robust}—in terms of watermark extraction bit error rate under cropping attacks of three sizes (64×64, 128×128, and 256×256) at three common cropping positions (upper-left, center, and upper-right), as well as the total runtime of the three schemes. Among them, Xiao et al.\cite{xiao2025robust} and Tang et al.\cite{tang2024robust} lack redundancy design in watermark embedding, making them non-robust to cropping attacks. The low bit error rate of Tang et al.\cite{tang2024robust} under the 64×64 cropping size is because the cropping did not affect the watermark embedding region. Wang et al.\cite{wang2024robust} exhibits certain robustness to cropping attacks, as it embeds complete watermarks independently in multiple circular domains. However, experimental results show that Wang et al.\cite{wang2024robust} has strong robustness against edge cropping attacks but poor robustness against center cropping attacks. This is because its design, which selects multiple feature points and forms embedding circular domains centered around these points to meet embedding capacity requirements, concentrates the embedding regions in the center of the image. The proposed scheme, by spirally embedding three copies of the watermark in dual-MSBs, ensures that the watermark is distributed across the center, edges, and intermediate transition regions of the image. Additionally, through a majority voting system, the receiver can correctly extract the uncropped watermark shares. As a result, the proposed scheme achieves the lowest watermark extraction bit error rate under all three cropping sizes (64×64, 128×128, and 256×256) and at different cropping positions. Moreover, for cropping sizes of 64×64 and 128×128, the proposed scheme enables completely accurate watermark extraction under cropping attacks.

\begin{table*}[!t]
\caption{Performance Comparison of Different Methods on BOSSbase Dataset}
\label{fixed crop}
\centering
\begin{tabular}{lcccccc}
\toprule
\textbf{Method} & \textbf{Crop Size} & \textbf{Top Left} & \textbf{Middle} & \textbf{Top Right} & \textbf{Average} & \textbf{Total Time (h)} \\
\midrule
\multirow{3}{*}{Tang et al.\cite{tang2024robust}} 
& 64$\times$64 & 0 & 0.172 & 0 & 0.057 & \multirow{3}{*}{37.696} \\
& 128$\times$128 & 39.430 & 2.711 & 38.610 & 26.917 & \\
& 256$\times$256 & 18.094 & 33.430 & 19.000 & 23.508 & \\
\addlinespace

\multirow{3}{*}{Wang et al.\cite{wang2024robust}} 
& 64$\times$64 & 0 & 2.454 & 0 & 0.818 & \multirow{3}{*}{24.895} \\
& 128$\times$128 & 0.474 & 10.512 & 0.260 & 3.749 & \\
& 256$\times$256 & 8.855 & 38.803 & 8.807 & 18.822 & \\
\addlinespace

\multirow{3}{*}{Xiao et al\cite{xiao2025robust}} 
& 64$\times$64 & 21.313 & 10.148 & 14.547 & 15.336 & \multirow{3}{*}{4.770} \\
& 128$\times$128 & 31.438 & 17.859 & 28.102 & 25.800 & \\
& 256$\times$256 & 46.453 & 28.750 & 45.805 & 40.336 & \\
\addlinespace

\multirow{3}{*}{Proposed} 
& 64$\times$64 & 0 & 0 & 0 & 0 & \multirow{3}{*}{5.622} \\
& 128$\times$128 & 0 & 0 & 0 & 0 & \\
& 256$\times$256 & 0.828 & 7.219 & 6.156 & 4.734 & \\
\bottomrule
\end{tabular}
\end{table*}

\begin{figure}[H]
\centering
\includegraphics[ width=0.3\textwidth]{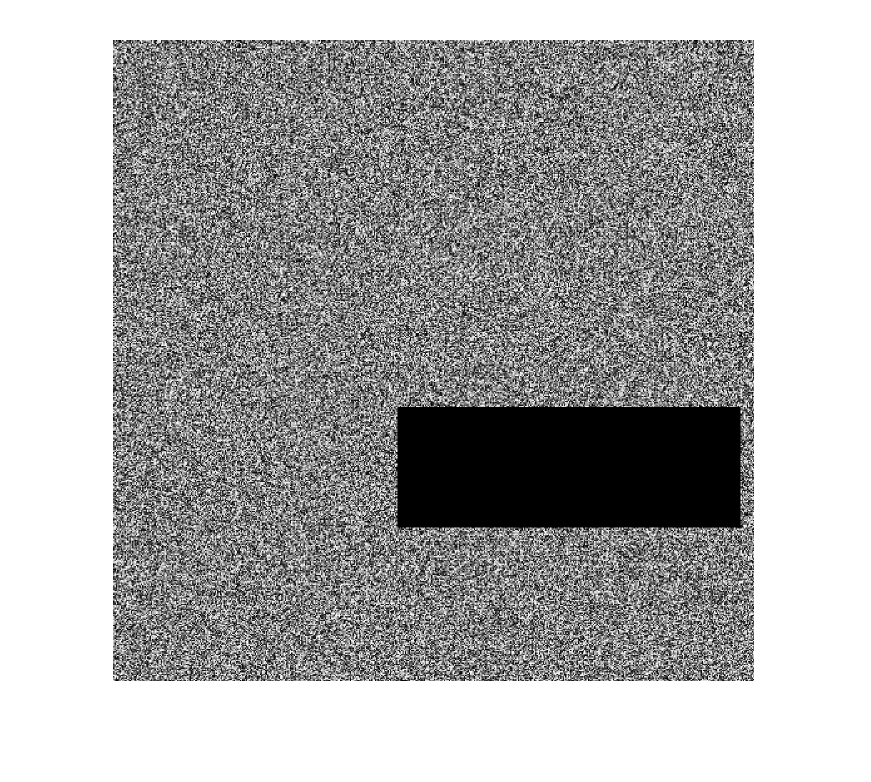} 
\caption{Cropped Encrypted image with 10\% cropping rate}
\label{rancrop1}
\end{figure}

\begin{figure}[htbp]
\centering
\includegraphics[trim = 0mm 80mm 0mm 80mm, clip, width=0.4\textwidth]{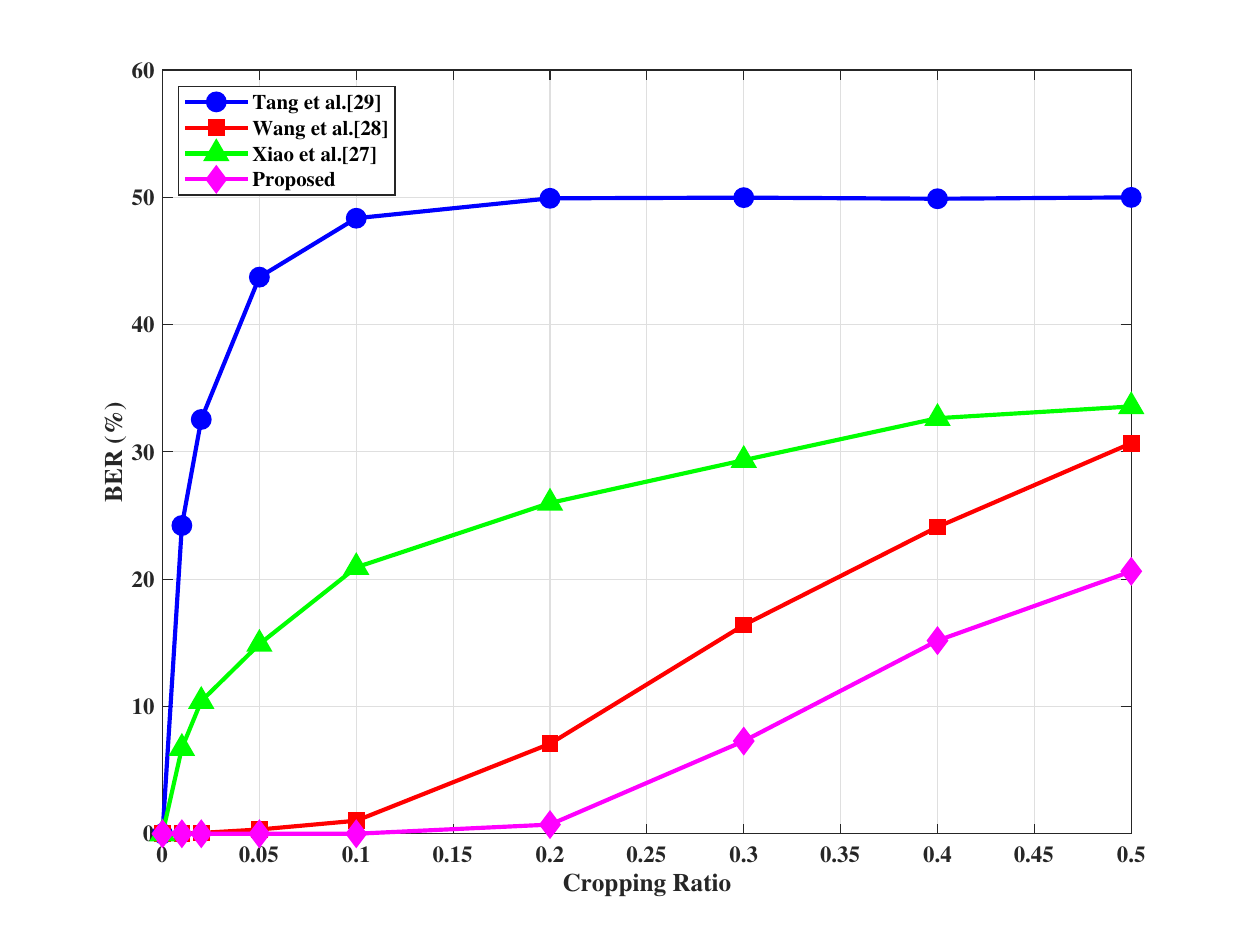} 
\caption{BER Performance Under Random Crop Attack on BOSSbase Dataset(when EC=128bits)}
\label{rancrop}
\end{figure}

Fig.\ref{rancrop} presents the cropping attack test conducted on 100 randomly selected images from the BOSSbase dataset. As shown in Fig.\ref{rancrop1}, the test involves varying ratios of cropped pixels to total pixels, with randomly generated aspect ratios and positions for the cropped regions. As shown in the Fig.\ref{rancrop}, Tang et al.\cite{tang2024robust} and Xiao et al.\cite{xiao2025robust} lack robustness against cropping attacks. In contrast, both Wang et al.\cite{wang2024robust} and the proposed method demonstrate robustness to cropping attacks due to their redundant watermark design. Furthermore, the proposed method exhibits superior robustness compared to Wang et al.\cite{wang2024robust}, maintaining accurate watermark extraction at the receiver end even when the ratio of cropped pixels reaches 10\%.

\begin{figure}[htbp]
\centering
\includegraphics[width=0.45\textwidth]{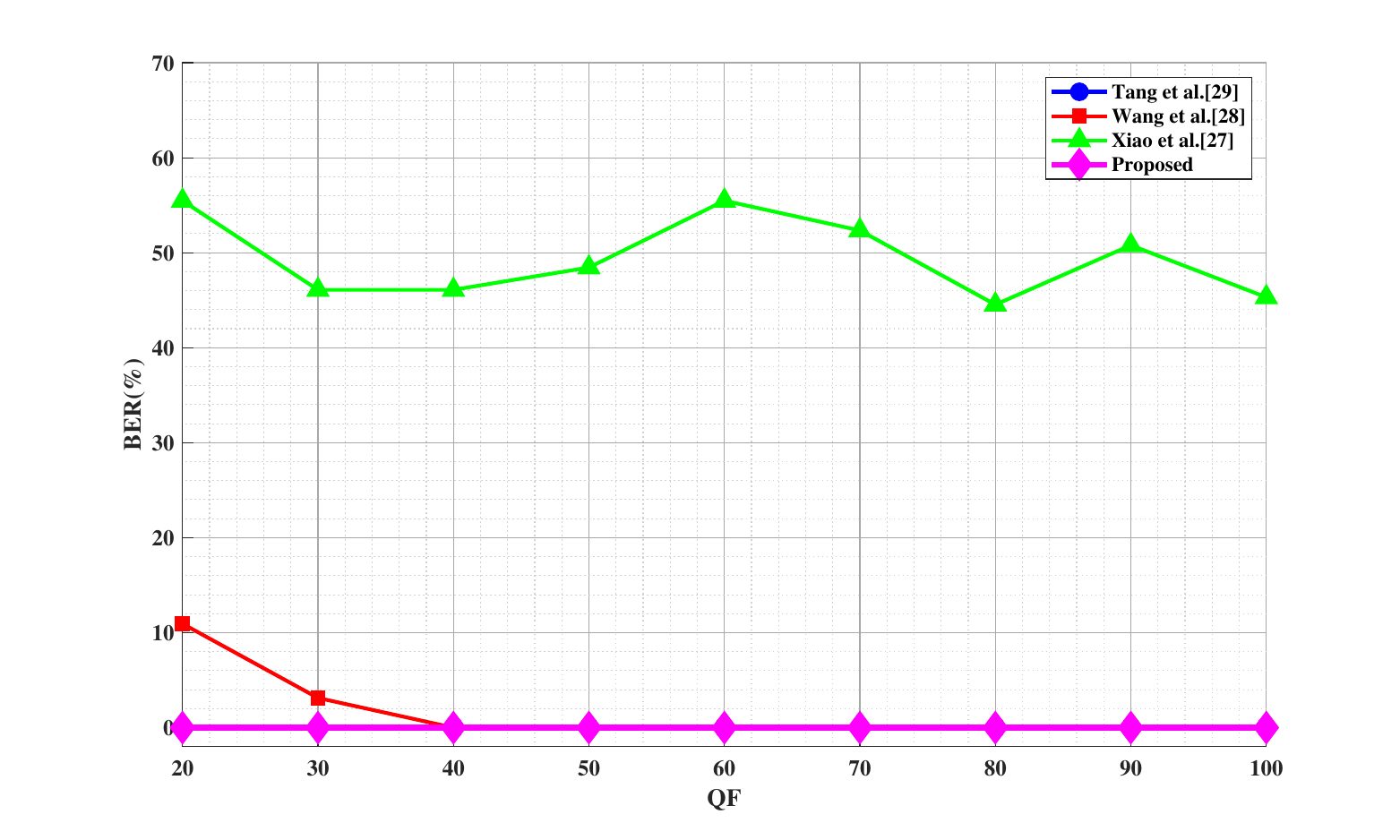} 
\caption{BER Performance Under JPEG compression Attack on BOSSbase Dataset(when EC=128bits)}
\label{JPEG}
\end{figure}

Fig.\ref{JPEG} reports the robustness evaluation of the embedded watermark under lossy compression, aiming to examine whether the watermark can be reliably extracted while preserving reversibility under common lossy operations encountered in practical storage, social media transmission, and image processing pipelines. JPEG compression, which is the most representative form of lossy compression, is adopted as the attack model. The JPEG quality factor (QF) is varied from 20 to 100, following common practice in the robust watermarking literature. This range deliberately excludes extremely low QF (QF $<$ 20), which lead to severe visual degradation and thus fail to provide meaningful evaluation in practical scenarios. Instead, it allows us to simulate a wide range of compression conditions, from relatively low to high quality, that are more likely to occur in real-world applications. A lower quality factor corresponds to a higher compression ratio. Experimental results demonstrate that the proposed method enhanced robustness within practical attack ranges against lossy compression. Similar to the Tang et al.\cite{tang2024robust} scheme, it achieves error-free watermark extraction throughout the range of QF = 20–100, and overall outperforms the Wang et al.\cite{wang2024robust} and Xiao et al.\cite{xiao2025robust} schemes.

\subsection{Statistical Security and Detectability Analysis}
\label{sec:stat_security}
The dual-MSBs strategy introduces a specific correlation between the dual-MSBs, thereby necessitating an analysis of the statistical security and detectability of the proposed scheme. In an ideal ciphertext image, bit‑planes should be mutually independent and possess maximum entropy to effectively resist statistical analysis attacks. However, to enhance robustness against noise, our method deliberately enforces the constraint $b7=b8$ to widen the effective noise margin. This constraint inevitably alters the joint probability distribution of the dual-MSBs, deviating from the ideal independent state.

\begin{figure}[htbp]
\centering

\begin{subfigure}[t]{0.3\linewidth}
    \centering
    \includegraphics[width=\linewidth, trim=0 60pt 0 0, clip]{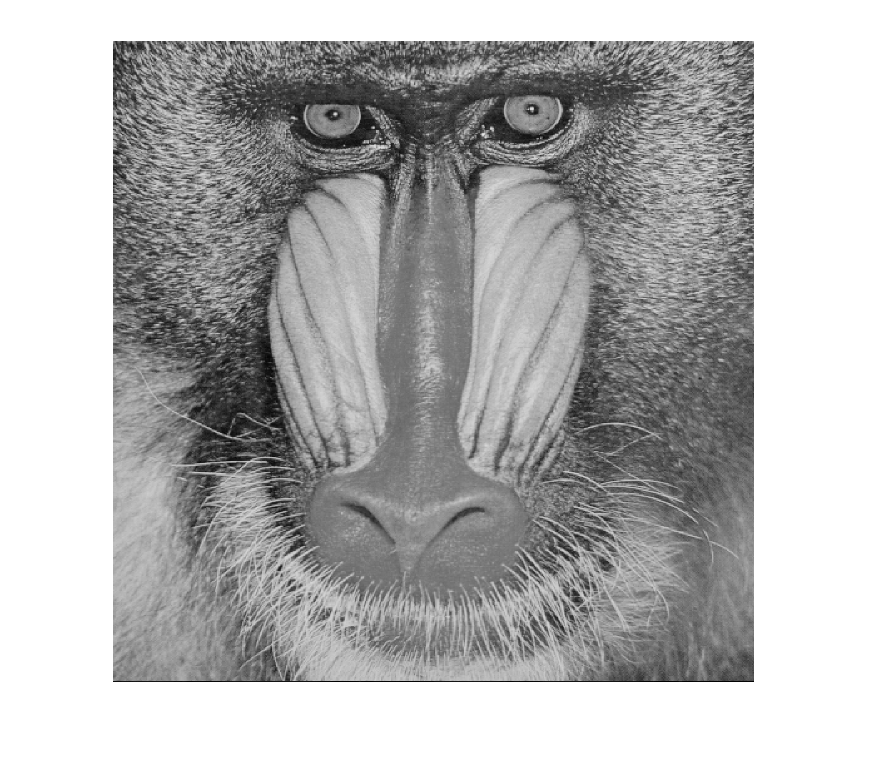}
    \caption{Baboon}
\end{subfigure}
\hfill
\begin{subfigure}[t]{0.3\linewidth}
    \centering
    \includegraphics[width=\linewidth, trim=0 60pt 0 0, clip]{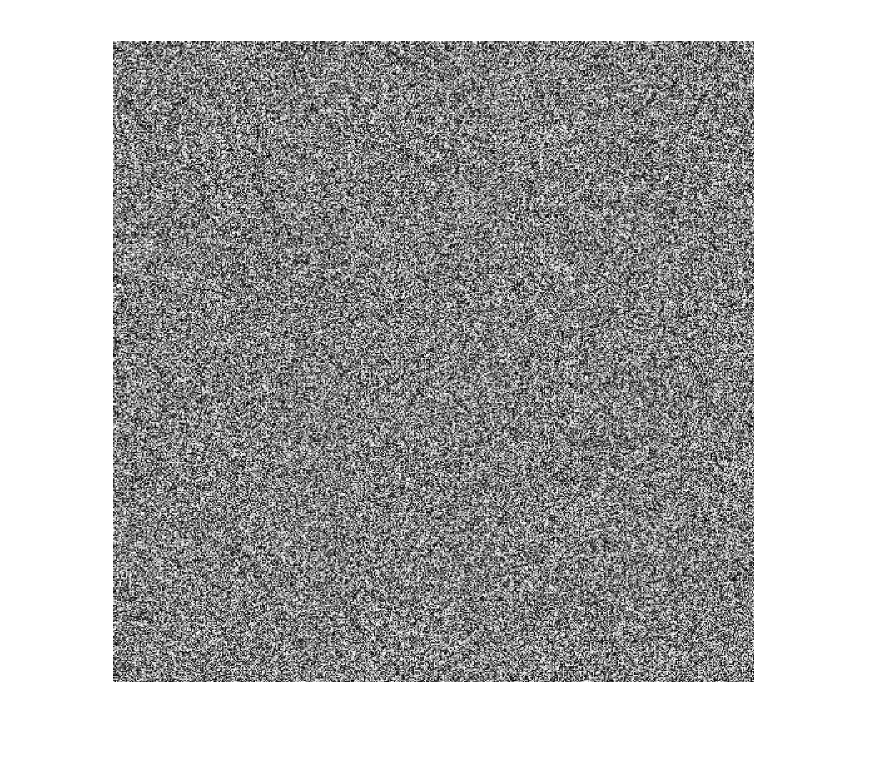}
    \caption{Encrypted image}
\end{subfigure}
\hfill
\begin{subfigure}[t]{0.3\linewidth}
    \centering
    \includegraphics[width=\linewidth, trim=0 60pt 0 0, clip]{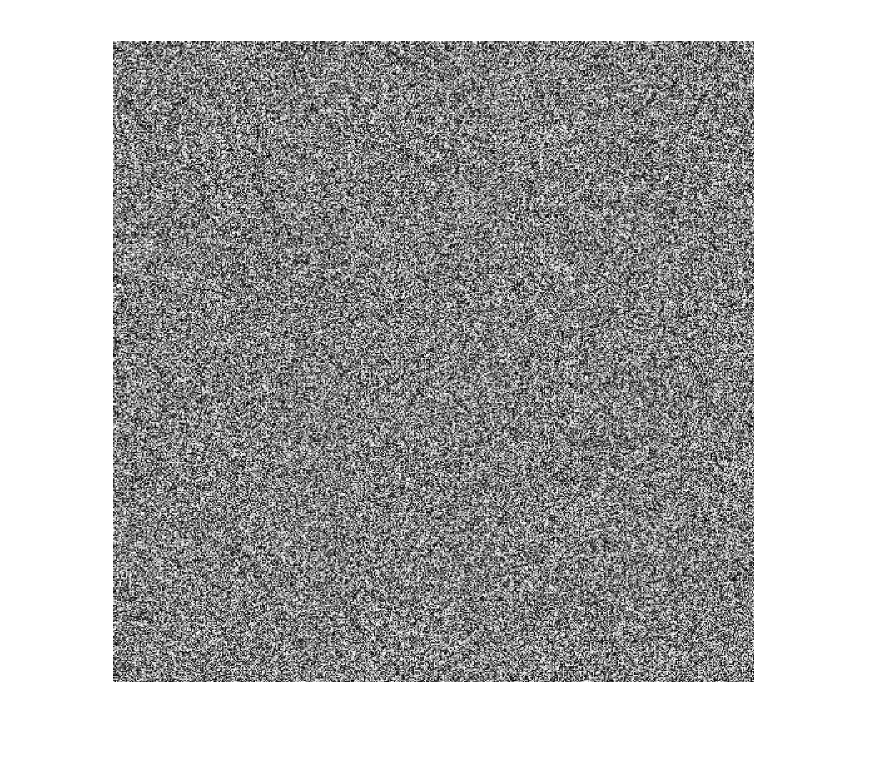}
    \caption{Watermarked image}
\end{subfigure}

\medskip

\begin{subfigure}[t]{0.3\linewidth}
    \centering
    \includegraphics[width=\linewidth]{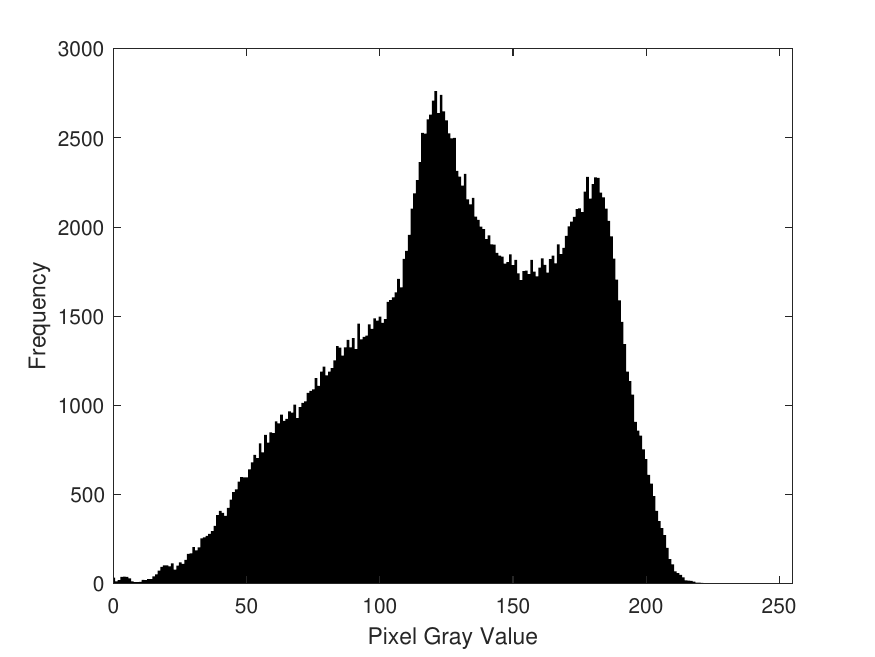}
    \caption{Histogram of original image}
\end{subfigure}
\hfill
\begin{subfigure}[t]{0.3\linewidth}
    \centering
    \includegraphics[width=\linewidth]{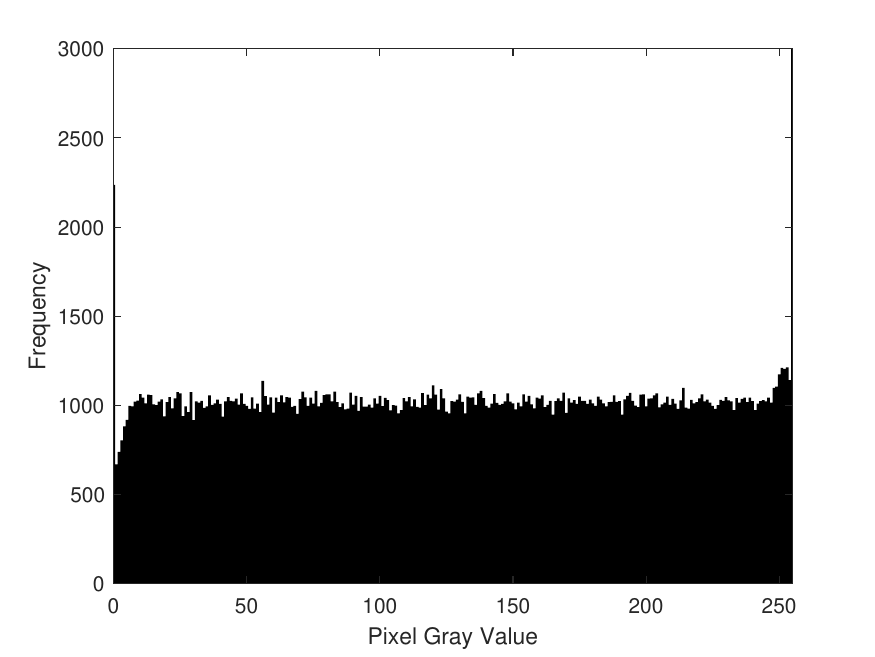}
    \caption{Histogram of encrypted image}
\end{subfigure}
\hfill
\begin{subfigure}[t]{0.3\linewidth}
    \centering
    \includegraphics[width=\linewidth]{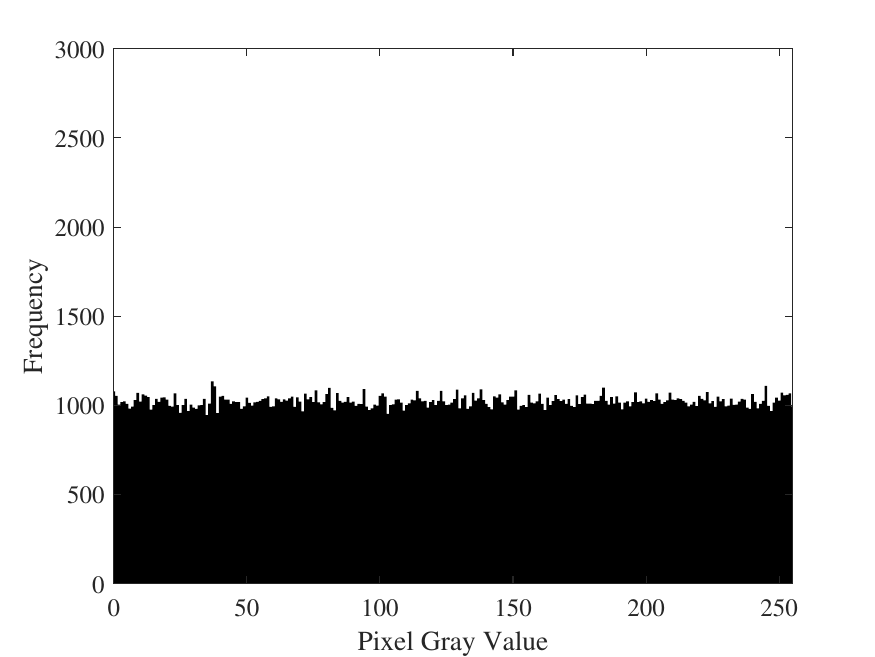}
    \caption{Histogram of watermarked image}
\end{subfigure}

\caption{The statistical characteristics of the Baboon image and its encrypted version: (a) original image; (b) encrypted image; (c) watermarked image; (d) histogram of original image; (e) histogram of encrypted image; (f) histogram of watermarked image.}
\label{fig:baboon-all}
\end{figure}

Nevertheless, comprehensive statistical evaluation suggests that the overall statistical security of the encrypted image and the final watermarked image remains high. As shown in Table \ref{tablesta}, the entropy of both the encrypted and watermarked images is close to 8, indicating that the pixel values are nearly uniformly distributed, which effectively conceals potential visual information. Moreover, the correlation coefficients in the vertical, horizontal, and diagonal directions are all close to 0, demonstrating that the encryption process successfully destroys the spatial correlation inherent in natural images. The NPCR between the encrypted/watermarked images and the original image reaches 99.60\%, indicating the high diffusion sensitivity of the encryption scheme. Additionally, as illustrated in Fig. \ref{fig:baboon-all}, the histograms of the encrypted and watermarked images become flat and uniform after encryption and subsequent watermark embedding, visually confirming that the pixel distribution does not reveal obvious statistical artifacts under first-order histogram and correlation analysis.

In summary, although dual‑MSBs embedding intentionally creates a local dependency between the two highest bit‑planes for robustness, the global statistical properties of the entire image, including entropy, pixel correlation, and histogram uniformity, are rigorously maintained at a level comparable to that of a secure ciphertext. This suggests that, under the evaluated statistical metrics, the proposed scheme does not introduce obvious statistical anomalies that could compromise content confidentiality, thereby achieving a favorable balance between robustness and statistical security.

\begin{table}[htbp]
\centering
\caption{Statistical Security Analysis of the Baboon Image}
\label{tab:correlation_npcr}
\begin{tabular}{c c c c c c}
\toprule
\multirow{2}{*}{Image} & \multirow{2}{*}{Entropy} & \multicolumn{3}{c}{Correlation} & \multirow{2}{*}{NPCR (\%)} \\
\cmidrule{3-5}
 & & V & H & D & \\
\midrule
Original image   & 7.3583 & 0.7587 & 0.8665 & 0.7262 & - \\
Encrypted image  & 7.9940 & 0.0002 & 0.0012 & -0.0011 & 0.9962 \\
Watermarked image & 7.9940 & -0.0002 & 0.0040 & 0.0027 & 0.9960 \\
\bottomrule
\end{tabular}
\label{tablesta}
\end{table}

\section{conclusion}
This paper presented a RRWEI framework that achieves strong robustness, high reversibility, and flexible embedding capacity at the same time. This method integrates techniques including prediction-error bit-plane compression, dual-MSBs embedding, spiral rearrangement, and Reed-Solomon code error correction. By meticulously designing the arrangement position and method of the watermark within the ciphertext-domain bit planes, it enhances the watermark's robustness against various attacks. The designed dual-MSBs embedding mechanism consistently outperforms under evaluated settings noise resistance, while the spiral embedding strategy ensures the recoverability of the watermark information under various cropping attacks. Based on extensive experiments conducted on the BOSSbase dataset, the proposed scheme demonstrates consistently better performance over existing robust reversible watermarking methods for encrypted images, as well as those robust reversible watermarking methods migrated from the plaintext domain to the ciphertext domain, in terms of noise resistance. It achieves a near-zero bit error rate under low to moderate noise levels and enables reliable watermark extraction even when confronted with large-area or randomly distributed cropping attacks. Future work may explore adaptive selection of the number of embedding copies and more robust distributions of the watermark within the bit planes against cropping attacks.
\bibliographystyle{IEEEtran}
\bibliography{references}

\newpage

\vspace{11pt}

\vspace{11pt}

\vfill

\end{document}